# THIN-FILM SCANDIUM ALUMINUM NITRIDE BULK ACOUSTIC RESONATOR WITH HIGH Q OF 208 AND K2 OF 9.5% AT 12.5 GHZ

*Sinwoo Cho[1], Yinan Wang[1], Eugene Kwon[2], Lezli Matto[2], Omar Barrera[1], Michael Liao[2], Jack Kramer[1], Tzu-Hsuan Hsu[1], Vakhtang Chulukhadze[1], Ian Anderson[1], Mark Goorsky[2], and Ruochen Lu[1]*

[1]The University of Texas at Austin, US, and [2]The University of California, Los Angeles

## ABSTRACT

This work describes sputtered scandium aluminum nitride (ScAlN) thin-film bulk acoustic resonators (FBAR) at 12.5 GHz with high electromechanical coupling ($k^2$) of 9.5% and quality factor ($Q$) of 208, resulting in a figure of merit (FoM, $Q \cdot k^2$) of 19.8. ScAlN resonators employ a stack of 90 nm thick 20% Sc doping ScAlN piezoelectric film on the floating bottom 38 nm thick platinum (Pt) electrode to achieve low losses and high coupling toward centimeter wave (cmWave) frequency band operation. Three fabricated and FBARs are reported, show promising prospects of ScAlN-Pt stack towards cmWave front-end filters.

## INTRODUCTION

The growing need for compact radio frequency (RF) filters in consumer electronics has spurred interest in acoustic wave technology [1], [2]. Traditional electromagnetic components face size limitations due to their longer wavelengths, but acoustic devices overcome this by converting signals to mechanical vibrations via piezoelectric materials [3], [4]. These vibrations operate at micron-scale wavelengths, enabling miniaturization while maintaining low signal loss and high-quality factors ($Q$) for precise frequency control [5]. Compatible with semiconductor/MEMS fabrication, acoustic resonators allow cost-effective mass production, making them vital for integrated RF systems like modern mobile transceivers that prioritize performance, size, and affordability [6], [7].

FR3 is a proposed frequency range, often referred to as the "upper-midband," that spans from 7.125 GHz to 24.25 GHz and sits between the FR1 (sub-6 GHz) and FR2 (millimeter-wave) bands used in current 5G networks [8], [9]. FR3 offers wider bandwidths and potentially better propagation characteristics compared to FR2, making it a promising option for cost-effective, high-performance future networks [10], [11]. In addition, the Ku-band (12–18 GHz), which belongs to FR3, balances spectral efficiency and coverage, making it ideal for 6G midbands [12], [13]. Scaling acoustic technologies, like thin-film aluminum nitride (AlN), is promising due to higher $Q$ and enhanced frequency selectivity, enabling low-phase-noise frequency references and high-performance filters with steep roll-off and rejection, building on its success in sub-6 GHz applications [14], [15]. Consequently, microwave acoustics provide high-performance signal processing functions, such as filters, in a compact form factor ideal for highly integrated RF systems, such as modern mobile transceivers with multiplexed RF paths sharing a single antenna [16], [17]. These capabilities outperform alternatives when simultaneously considering size, power, performance, and cost [18].

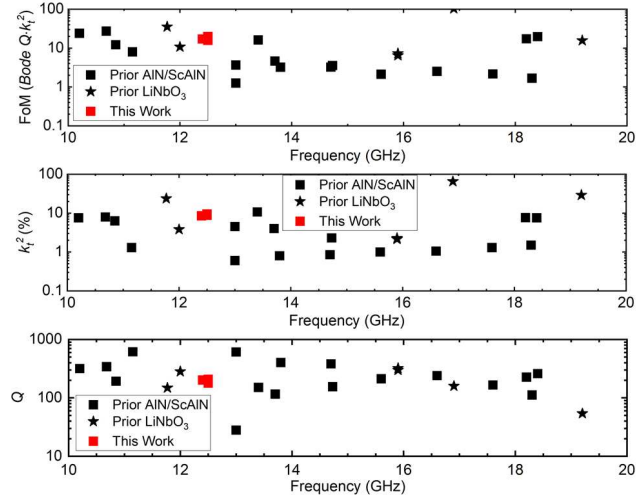

*Fig. 1 Survey of reported resonators above 10 GHz (a) FoM (b) $k_t^2$ (c) Q.*

However, earlier efforts often faced difficulties in efficiently using acoustics over a wide bandwidth while achieving high performance, due to fundamental limitations like low electromechanical coupling ($k^2$) and high damping in traditional piezoelectric materials [19], [20]. Recent advancements in materials, design, and fabrication have shown that thin-film scandium aluminum nitride (ScAlN) film bulk acoustic resonators (FBARs) are promising to achieve higher $k_t^2$ [21], [22]. In recent years, scandium doping has been introduced as a method to enhance the piezoelectric coefficients and material compliance of AlN [23]. To be specific, highly doped ScAlN (with Sc concentrations of 20% or more) has emerged as a promising material for FBARs [24], [25]. Thus, it makes ScAlN a preferred choice for many modern RF and MEMS devices compared to traditional AlN [26], [27]. As a result, they offer low loss, wide bandwidth, and frequency diversity, making them suitable for high-performance 5G front-end signal processing applications [28], [29].

However, previous research indicates the difficulty of scaling ScAlN/AlN FBARs, as demonstrated by the presentation of moderate figures of merit (FoM, $k^2 \cdot Q$) at Ku-bands (Fig. 1) [30]. First, FBAR requires a film stack of sub-100 nm thick piezoelectric and sub-50 nm electrode layers at Ku-bands [31]. Keeping good crystalline during sputtering is challenging due to surface roughness and oxygen level in intermediate boundaries [32]. Secondly, few studies analyze the impact of different electrode

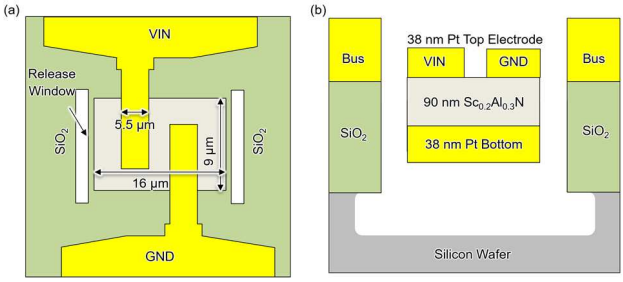

*Fig. 2 (a) Top and (b) cross-sectional view of ScAlN FBAR.*

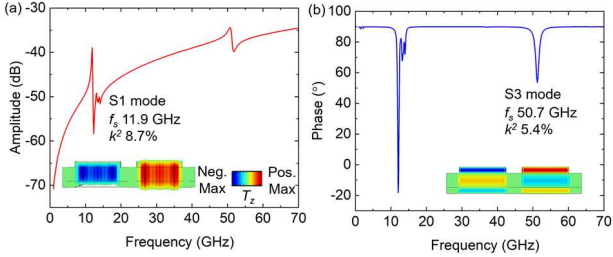

*Fig. 3 FEA simulated wideband admittance (a) amplitude and (b) phase with simulated displacement mode shapes at S1 and S3 resonance.*

materials on $k^2$ and $Q$ [33]. The effects of electrodes are essential, especially in terms of acoustic impedance, stiffness, and ability to help orient piezoelectric films in determining suitability of a given material to be used for a specific FBARs application [34]. For most applications, it is ideal for acoustic waves traveling between two electrodes to have the majority of their energy reflected by the electrode [35], [36]. The extent to which the wave penetrates into the electrode before being reflected depends on the material's acoustic impedance and stiffness [37]. Materials with high acoustic stiffness reflect more energy with minimal wave penetration, resulting in minimal energy loss [38]. All of the above challenges require innovative design and fabrication toward Ku-Band FBARs [39].

In this work, we report the influence of Pt metal electrodes on sputtered ScAlN FBARs, which can be a great potential electrode platform, dramatically increasing $k_t^2$ and $Q$ at Ku-band. The demonstrated FBAR with Pt top and bottom electrodes achieved $k^2$ of 9.5% and $Q$ of 208 for the first-order symmetric (S1) mode at 12.5 GHz, resulting in a FoM of 19.8. Through these results, we confirmed that FBARs acoustic filter at Ku-band may be achievable.

## DESIGN AND SIMULATION

Figure 2 (a)-(b) shows the FBAR's top and cross-sectional structure. The design features a bottom floating electrode and top electrodes (ground/signal traces), with a 90 nm Sc-doped AlN ($Sc_{0.2}Al_{0.8}N$) layer sandwiched between 38 nm Pt electrodes. The thin film's high capacitance density enables compact resonator dimensions (16 × 9 μm). Low-resistance aluminum interconnects (300 nm thick) minimize routing losses, while a lateral $SiO_2$ layer encircling the resonator suppresses feedthrough-induced parasitic resistance and capacitance [40]. The proposed FBAR [Fig. 3 (a)-(b)] is simulated using COMSOL finite element analysis (FEA) with a mechanical $Q$ value of 50. In operation, the electric field between top and bottom electrodes excites a confined thickness extensional mode in ScAlN. The piezoelectric transduction is achieved mostly from thickness stress component [$T_z$ in Fig. 3 (a)-(b)], via piezoelectric coefficient $e_{33}$. The ScAlN thickness, and Pt thickness are selected with consideration of high $k^2$ at 11.9 GHz and spurious free resonance result. The simulated results show $k^2$ of 11.9% and 5.4% for S1 mode at 11.9 GHz and S3 mode at 50.7 GHz, respectively. The extracted each $k^2$ from S1 and S3 mode, obtained via Butterworth-Van Dyke (BVD) fitting, which is equivalent to

$$k^2 = \pi^2/8 \cdot (\frac{f_p^2}{f_s^2} - 1)$$

for this case, without EM effects, e.g., routing inductance and resistance [41]. Thus, it is expected that the appropriate stack for FBAR predicts that device performance can be maximized.

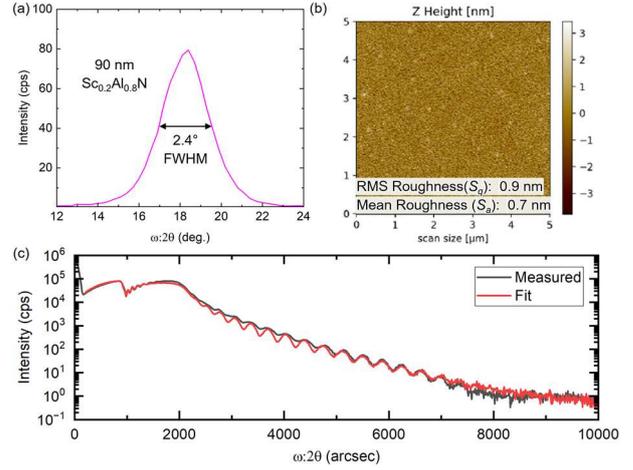

*Fig. 4 (a) XRD Rocking curve of the ScAlN layer, (b) ScAlN film surface measurement by AFM, and (c) XRR measurement of the stack.*

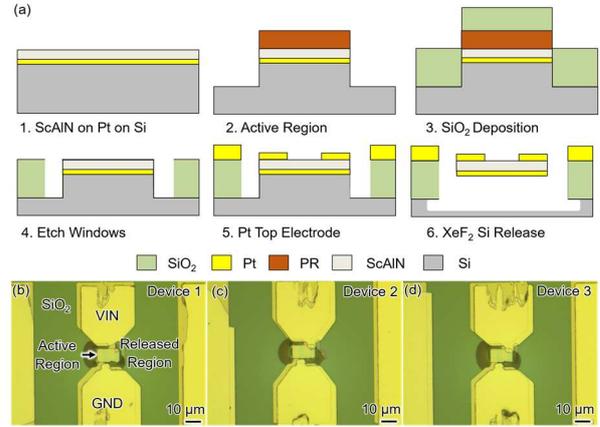

*Fig 5: (a) FBAR fabrication flow chart and their images of (b) device 1, (c) device 2, and (d) device 3.*

## MATERIAL ANALYSIS & FABRICATION

The fabrication process starts with depositing a 38 nm layer of Pt followed by a 90 nm layer of ScAlN onto a high-resistivity silicon (Si) <100> wafer (> 10,000 Ω·cm). This deposition is carried out using the Evatec Clusterline 200 sputtering system, which maintains a continuous vacuum environment. Quantitative material analysis begins with X-ray diffraction (XRD) for each layer stack, as shown in Fig.

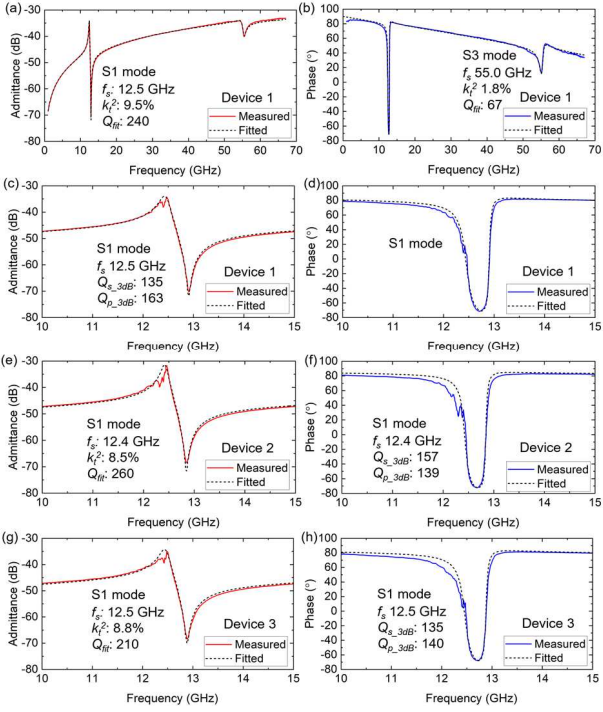

*Fig. 6 Measured admittance in amplitude and phase of device 1 (a)-(d), device 2 (e)-(f), and device 3 (g)-(h).*

4 (a) and (c). The rocking curve of the $Sc_{0.2}Al_{0.8}N$ layer on the bottom Pt electrode shows a full width at half maximum (FWHM) of 2.4°, indicating improved crystal quality in the sputtered thin film [42]. Figure 4(b) presents the surface roughness of the ScAlN layer as measured by atomic force microscopy (AFM) [43].

Figure 5 (a) outlines the fabrication steps. The process begins by etching the ScAlN, Pt, and Si layers around the active area using the AJA Ion Mill [44]. A $SiO_2$ layer is then deposited over these etched regions via low-temperature (100 °C) plasma-enhanced chemical vapor deposition (PECVD), ensuring electrical insulation and preventing disconnection of the top electrode due to step height differences. The AJA Ion Mill is then used again to define and etch release windows. Following this, aluminum buslines, contact pads, and top electrodes are deposited using a KJL e-beam evaporator. The final structural release is achieved through isotropic silicon etching with xenon difluoride ($XeF_2$). Optical images of the fabricated FBAR structures are shown in Fig. 5 (b)–(d), all sharing identical designs and dimensions.

## MEASUREMENT AND DISCUSSION

The resonators are characterized using a Keysight vector network analyzer (VNA) in a two-port configuration at room temperature in air, with an input power of −15 dBm. Figures 6 (a)–(h) show the measured admittance magnitude and phase. The wideband admittance responses in both amplitude and phase for the S1 mode of each FBAR are presented in Fig. 6 (c)-(d), (e)-(f), and (g)-(h), respectively. A modified Butterworth–Van Dyke (mBVD) model is applied to extract the performance parameters of each resonator, incorporating series resistance ($R_s$) and inductance ($L_s$) to account for electromagnetic (EM) effects [45]. Prior to adding motional components to extract quality factor ($Q$) and electromechanical coupling ($k^2$), the EM-related parameters $R_s$, $L_s$, and the static capacitance $C_0$ are first fitted using the admittance magnitude and phase.

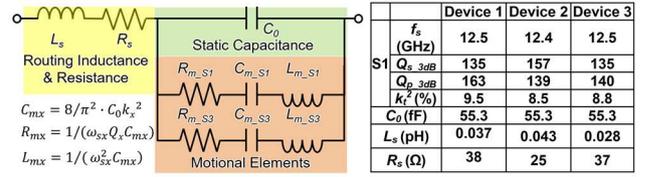

*Fig 7: Modified MBVD model and extracted key resonator specifications.*

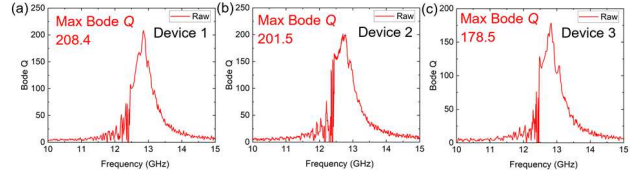

*Fig. 8 Each FBAR Bode Q of S1.*

The resulting fitted curves of FBARs (device 1-3) are shown in Fig. 6 (a)–(h). The device 1 FBAR in Fig.6 (c)-(d) achieves $k^2$ of 9.5% and $Q_{fit}$ of 240 for S1 mode. The FBAR from device 2, shown in Fig. 6 (e)-(f), demonstrates a $k^2$ of 8.5% and $Q_{fit}$ of 260 for S1 mode. The FBAR from device 3, depicted in Fig.6 (g)-(h), exhibits a $k^2$ of 8.8% and $Q_{fit}$ of 210 for S1 mode. These obtained parameters are detailed in Fig. 7 table, and show consistency of good FBARs performance with high $Q_{fit}$ and $k^2$. To further verify this, Fig. 8 (a)-(c) exhibits the Bode $Q$ values for each device's S1, respectively which are 208.4, 201.5, and 178.5 for them.

Its high FoM at S1 mode shows the possibility of proper usage of metal electrodes might achieve good performance of resonator with enhanced $Q$ and $k^2$ at cmWave. Our study exhibits higher FoM to previous ScAlN/AlN works for the Ku-band. In addition, three fabricated FBARs are reported, showing the reliability and consistency of device design and performance. Their promising prospects of ScAlN-Pt stack towards cmWave front-end filters are realizable.

## CONCLUSION

This study presents sputtered ScAlN thin-film FBARs operating at 12.5 GHz, achieving a $k^2$ of 9.5% and a $Q$ of 208, which together yield FoM of 19.8. We notice Pt electrode FBARs show higher $Q$ and $k^2$ coming from high acoustic impedance, stiffness, and ability to help orient piezoelectric films in determining suitability of a given material to be used for FBAR applications. These results confirmed that even in the frequency band of cmWave, ScAlN FBAR can achieve enough performance with optimized fabrication and acoustic/EM design for FBAR filters.

## ACKNOWLEDGEMENTS

The authors would like to thank the funding support from the DARPA COFFEE program, Dr. Ben Griffin, Dr. Todd Bauer, and Dr. Zachary Fishman for the helpful discussion.

# CONTACT

sinwoocho@utexas.edu